\documentstyle[prd,aps]{revtex}
\begin{document}
\input epsf \renewcommand{\topfraction}{0.8}

\catcode`@=11
\def\gsim{\mathrel{\mathpalette\@versim>}}
\def\@versim#1#2{\lower0.2ex\vbox{\baselineskip\z@skip\lineskip\z@skip
      \lineskiplimit\z@\ialign{$\m@th#1\hfil##$\crcr#2\crcr\sim\crcr}}}
\catcode`@=12
\catcode`@=11
\def\lsim{\mathrel{\mathpalette\@versim<}}
\def\@versim#1#2{\lower0.2ex\vbox{\baselineskip\z@skip\lineskip\z@skip
       \lineskiplimit\z@\ialign{$\m@th#1\hfil##$\crcr#2\crcr\sim\crcr}}}
\catcode`@=12
\draft
\twocolumn[\hsize\textwidth\columnwidth\hsize\csname
@twocolumnfalse\endcsname

\title{Classification of Inflationary Potentials}
\author{\sc Mikel Susperregi}\footnote{m.susperregi@qmw.ac.uk}\\
\\
\address{Astronomy Unit, School of Mathematical Sciences, 
Queen Mary \& Westfield College, \\
University of London, London E1 4NS, United Kingdom}
\date{\today}
\maketitle
\begin{abstract}
Brans-Dicke gravity is remarkable not only in that General 
Relativity and Mach's Principle find a common enlarged scenario where 
they are mutually consistent, but also in that it provides a very 
interesting quantum cosmological model within the 
inflationary paradigm. The interplay between the Brans-Dicke scalar 
$\Phi$ and the inflaton field $\sigma$ plays an important r\^{o}le 
during the course of inflation, and although the dynamics as such is governed 
by the potential, the onset and the end of inflation are determined 
by the values of both fields jointly. The relative 
position of the beginning-- and end-of-inflation curves (BoI and EoI 
respectively) is the most relevant factor in determining the 
resulting quantum cosmological scenario. The classification of 
potentials that is given in this paper is based on the criterion of 
whether the BoI and EoI boundaries enclose a finite or infinite area 
in the ($\sigma$,$\Phi$) plane where inflation takes place. It is 
shown that this qualitative classification distinguishes 
two classes of potentials that yield very different cosmologies 
and it is argued that only those theories in which BoI and EoI 
enclose a finite area in the ($\sigma$,$\Phi$) plane are compatible 
with our observable universe. 
\end{abstract}


\vskip2pc]


\section*{}

One of the key features that makes the inflationary paradigm successful 
is that it provides a mechanism to explain the origin of fluctuations 
and predict a scale-invariant Harrison-Zel'dovich spectrum at the end 
of inflation \cite{spectrum,inflation}. Chaotic inflation
\cite{inflation,chaotic} in particular 
describes the universe as a statistical ensemble of regions, thus 
enabling us to envisage a quantum cosmological scenario whose
properties are entirely dependent on the potential of the scalar field. 
In these theories, homogeneous regions in an ensemble of universes are 
subdivided into further regions where the scalar field $\sigma$ takes 
a large number of values. The scalar field evolves according to a stochastic 
equation $\dot\sigma= \dot\sigma_S +\xi$ that describes its Brownian
motion, as formulated in \cite{stochastic}. The value of the field 
is hence the combination of the slow-roll solution $\sigma_S$, that is the 
solution of $\dot\sigma_S= -(3H)^{-1} V'$ and is due to  
the contribution of the fields over scales $\gsim H^{-1}$, 
and quantum fluctuations within $\lsim H^{-1}$ that are 
related to the stochastic term $\xi$. The Starobinsky noise $\xi$ follows 
a Gaussian distribution centered around the classical solution
$\sigma_S$ and its variance is $\sim H/2\pi$. 

An arbitrary region $\epsilon H^{-1}$ of an initial homogeneous volume 
$\sigma=\sigma_A$ that undergoes a quantum jump $\delta\sigma$ 
in a timestep $\delta t$ is taken to a new classical trajectory 
that corresponds to the hypersurface $\sigma=\sigma_A+\delta\sigma$, 
and the forthcoming fluctuations at the next timestep within 
this region are gaussianly 
distributed around the new classical value. The distribution of the 
so-called {\it coarse-grained} field is then governed by 
the Fokker-Planck equation and it is only dependent on the 
form of the potential \cite{stochastic,coarse,stationary}. 
The inflationary expansion of each region that results from this continuous 
subdivision of homogeneous regions due to quantum jumps depends 
on the local value of $\sigma$, and thus quantum fluctuations that 
take the scalar field to larger values expand much faster than 
those that take it to smaller values. After a sufficiently 
long span of inflation, quantum fluctuations take the scalar 
field step by step to larger and larger values, and the volume 
of the universe becomes dominated by regions with values of $\sigma$ 
close to the largest permissible for a given potential. 
This situation is summarized in the simplest terms in Fig.~1. 
The starting point in this example 
is a homogeneous bubble $\sigma=\sigma_0$, shown at the top of the 
figure. After a brief lapse of inflation $\delta t$, quantum 
fluctuations divide the region in two halves; these are homogeneous 
regions resulting from a positive quantum fluctuation (right, denoted 
$\uparrow$), and a negative one (left, $\downarrow$) 
around $\sigma=\sigma_0$. Both fluctuations are for simplicity 
of the same amplitude $\delta\sigma$, as will all subsequent 
fluctuations. The volume of the region $\uparrow$ on the right expands faster 
than the region $\downarrow$ on the left and is larger by a 
factor $\sim e^{6H^{\prime}\delta\sigma\delta t}$. 
After a second lapse of time $\delta t$, the two regions 
$\uparrow$ and $\downarrow$ undergo further divisions in 
very much the same way and yield four homogeneous regions 
from the subdivision of the previous two in positive and 
negative quantum fluctuations of the same amplitude. These volumes 
are labelled $\downarrow\downarrow$,$\downarrow\uparrow$,
$\uparrow\downarrow$,$\uparrow\uparrow$. The relative sizes of 
the bubbles in the figure are not to scale but are only intended to 
illustrate that some volumes are bigger than others (typically by 
many orders of magnitude). Certainly, in this representation where 
we have chosen identical fluctuation amplitudes $\delta\sigma$,  
the regions $\uparrow\downarrow$ and $\downarrow\uparrow$ are of 
equal size, so their volumes add up. A more realistic situation of 
the phenomenon depicted in the toy representation of Fig.~1 is 
that the quantum 
fluctuations $\delta\sigma$ follow a Gaussian distribution, and 
therefore the many resulting regions of the type 
$\uparrow\downarrow$, $\uparrow\downarrow\uparrow\downarrow$ are
really part of vast number of regions for which positive and negative 
fluctuations even out and therefore lay on the trajectory that follows
the classical evolution of $\sigma=\sigma_0$.

\begin{figure}[t]
\centering
\leavevmode\epsfysize=5.5cm \epsfbox{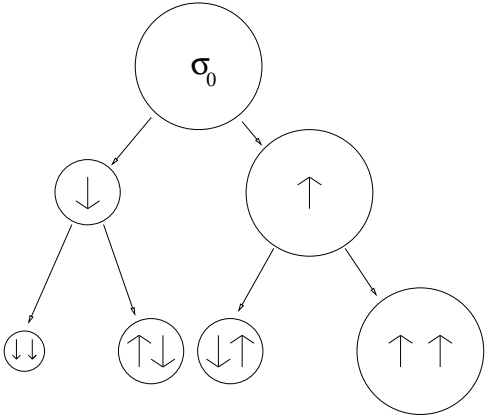}\\
\vskip 0.2cm
\caption[fig1]{Simple sketch of the division of homogeneous 
regions due to quantum jumps. Positive quantum 
jumps ($\uparrow$) are located to the right of each subdivision, 
whereas negative jumps ($\downarrow$) of the same magnitude 
are located to the left. The total volume is dominated at every stage 
by the regions with largest values of the fields.}
\end{figure}

\begin{figure}[t]
\centering
\leavevmode\epsfysize=5.5cm \epsfbox{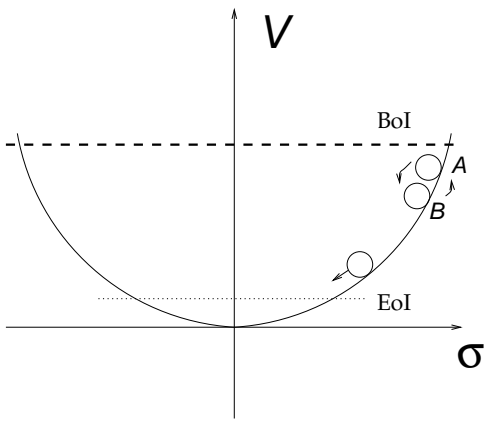}\\
\vskip 0.2cm
\caption[fig2]{Powerlaw potential. Beginning- (BoI) and 
end-of-inflation (EoI) boundaries are represented by the 
slashed and dotted straight lines respectively. BoI is 
given by $V(\sigma)\approx M_P^4$ whereas EoI is determined 
by $\dot\sigma^2\approx 2V(\sigma)$. The configuration at 
$A$ is sustained by quantum jumps from smaller values of $\sigma$, 
such as $B$.}  
\end{figure}

As was shown in \cite{centre}, following the argument sketched 
above, the largest volume of the universe is occupied by values of 
$\sigma$ that correspond to the maximum values of the potential, 
i.e. in the neighbourhood of the Planck or beginning-of-inflation 
boundary (BoI), $V(\sigma)\approx M_P^4$. It can be seen in 
Fig.~2, for the case of a powerlaw potential, that 
the Planck boundary sets a delimiter 
to the rate of expansion that corresponds to a certain 
$\sigma\approx \sigma_{\rm max}$ and the volume of the universe 
is dominated by regions where the scalar field takes this value. 
Whereas the field will classically slow-roll down the potential, 
quantum jumps that take $\sigma$ back up to larger values will 
be enhanced as they create volumes that grow faster, and therefore, 
as is shown in Fig.~2, the field that evolves classically from 
the initial location $A$ to $B$ is typically taken back up to 
$A$ via a quantum transition. Quantum fluctuations retain the field 
in the neighbourhood of BoI for as long as possible, and therefore  
an arbitrary region of the universe that has undergone thermalization 
is most likely to result from a region where the scalar 
field has stayed in the neighbourhood of BoI for a long time and 
then rolled down the potential as quickly as possible, via coherent 
quantum jumps, to cross the end-of-inflation (EoI) boundary. 

In this simplest case of inflation driven by one scalar field $\sigma$, 
the BoI is a scale inherent to the physics that depends on the 
value of $M_P$, which is a constant of Nature, and the amplitude 
of $V$ is tuned to predict the right order of magnitude in the 
spectrum of fluctuations, etc. On the other hand, scalar-tensor theories 
of gravity (see e.g. \cite{scalar-tensor} for a review) provide a means to
investigate $M_P$ in terms of a scalar field, e.g. the Brans-Dicke 
(BD) field $\Phi$, so that $M_P$ varies from one region to another in 
the universe like a coarse-grained field, in much the same way 
as the inflaton in the chaotic model; the combination of these 
$G$-varying theories with chaotic inflation 
yields interesting scenarios where the interplay between $\Phi$ and 
the inflaton $\sigma$ determines the distribution of $M_P$ in an 
ensemble of universes in terms of the potential of the scalar field, 
and therefore, ultimately in the particle physics involved. 
Two interesting scenarios that result from the combination of 
BD gravity and chaotic inflation are induced \cite{induced} and 
extended \cite{extended} inflation. The difference between both 
is that in the former the BD field has a non-vanishing potential, 
whereas the latter has not. For simplicity in this paper we will 
investigate extended inflation only, and the same analysis applies 
to induced inflation in the straightforward way. 

The extended inflation action is \cite{extended} 
\begin{equation}
\label{action}
S =\int d^{4}x \,\sqrt{-g}\left[\Phi R
   - {\omega\over\Phi}(\partial \Phi)^{2} 
 - \frac{1}{2}(\partial \sigma)^{2} \\
  - V(\sigma)\right] ,
\end{equation}
where $V(\sigma)$ is the inflation potential, $\Phi$ is the BD field, 
$\omega$ is the BD coupling, that we take to be constant. The
slow-roll variational equations in an FRW background take the 
form
\begin{eqnarray}
\label{SR}
\frac{\dot\Phi}{\Phi} &=&2\frac{H}{\omega} \,, \\
\dot \sigma &=& -\frac{1}{3H}V^{\prime}(\sigma) \,,\\ 
H^{2} &=&\frac{1}{6\Phi}V.
\end{eqnarray}
The BoI boundary is $V(\sigma)=M_P^4(\Phi)$ (where 
$M_P^2(\Phi)\equiv 16\pi\Phi$), and EoI is given by 
\begin{equation}
\frac{1}{2}\dot\sigma^2+\omega{\dot\Phi^2\over\Phi}\approx V(\sigma).
\end{equation}
With the aid of (\ref{SR}) EoI is rewritten as 
\begin{equation}
\label{EoI}
 \Phi = \left({3\omega 
-2\over \omega}\right)\left({V\over V^{\prime}}\right)^2.
\end{equation}
From (\ref{SR}) the following conservation law follows 
\cite{mikel1,mikel2} 
\begin{equation}
\label{integrals}
\frac{d}{dt}\left[ \omega\Phi+\int\! d\sigma{V(\sigma)
\over V^{\prime}(\sigma)}\right] =0,
\end{equation}
which emerges due to the simplicity of the slow-roll equations 
and does not correspond to a symmetry inherent in the physics. 

The functional form of BoI and EoI suggests the following
classification of $V$ in two classes, based on the relative 
position of the boundaries of inflation:
\\
\begin{itemize}
\item {\bf Class I} {\it The area of inflating regions enclosed 
between the boundaries BoI and EoI is finite.} In this case, 
the BoI and EoI curves intersect at non-trivial values 
of the fields.  
\\
\item {\bf Class II} {\it The area enclosed by BoI and EoI is 
infinite.}
\end{itemize} 

In the following we illustrate these classes with some examples, 
and argue that this qualitative classification distinguishes 
a very important feature that leads to two entirely different 
cosmological scenarios. 

Firstly we investigate powerlaw potentials $V(\sigma)=
\lambda/(2n)\,\sigma^{2n}$ of both {\it class I} and {\it class II}. 
Fig.~3 shows the BoI and EoI boundaries for a powerlaw potential 
for several $n$. The EoI boundary is 
a parabola for all $n$ (and only its amplitude is scaled by 
a factor $1/(4n^2)$). The BoI boundary on the other hand strongly 
depends on $n$, i.e. $\Phi\sim \sigma^n$. For $n=1,2$ BoI does not 
intersect EoI, and both boundaries span an infinite region on 
the ($\sigma$,$\Phi$) plane. Therefore $n=1,2$ are {\it class II} 
potentials, whereas $n>2$ are on the other hand {\it class I}. 
The classical trajectories (\ref{integrals}) are inverted parabolas and the 
fields move in the direction $A\to B$, in concentric parabolas, 
along the segment of these curves that is contained within BoI and 
EoI. Initially, the fields that start out at ($\sigma_0$,$\Phi_0$) 
move along the parabola that crosses this point, though  
quantum fluctuations soon disperse the motion of the fields 
to concentric trajectories. In turn, regions that have switched 
to other classical trajectories via quantum jumps will predominantly 
follow the slow-roll trajectory, and they are further subdivided 
in other domains due to later quantum fluctuations that take 
the fields to other adjacent classical trajectories. The end result, 
qualitatively, is that regions that move to larger and larger 
values of $\sigma$ dominate the total volume of the universe. 
The important point to note here, as opposed to the $\sigma$-only 
scenario sketched in Fig.~1 where BoI is a fixed quantum scale, 
is that potentials of the second class, such as $n=1,2$, allow 
$\sigma$ to grow indefinitely while remaining in the region enclosed 
by BoI and EoI. In this situation, any arbitrarily large value 
of the fields is much likelier and predominant than smaller values, 
but at the same time its likelihood is negligible in comparison 
to larger values. Hence, in the scenario described by potentials 
of {\it class II}, the fields have no typical values and the 
dominant contribution to the total volume is one where both 
fields blow up.\footnote{The introduction of a non-minimal coupling 
$\xi\sigma^2 R$ 
was suggested by \cite{nonminimal} to remedy this situation 
and transform {\it class II} powerlaw potentials into {\it class I}, 
although \cite{nonminimal} did not note that $n>2$ are in fact 
per se {\it class I} regardless of non-minimal coupling.} 

\begin{figure}[t]
\centering
\leavevmode\epsfysize=5.5cm \epsfbox{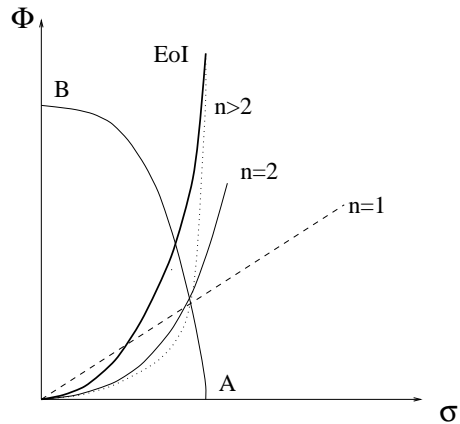}\\
\vskip 0.2cm
\caption[fig3]{BoI and EoI boundaries for powerlaw potentials. 
EoI is thick solid line, scaled to the same amplitude in 
all three cases $n=1,2$ and $n>2$. BoI is given by: slashed line ($n=1$); 
solid line ($n=2$); dotted line ($n>2$). Classical 
trajectories are parabolae concentric to $AB$. BoI and EoI intersect 
in the case of $n>2$.}
\end{figure}

In contrast to this scenario, potentials $n>2$ describe a very 
different universe. As it can be seen in Fig.~3, $n>2$ are 
potentials of the first class: BoI and EoI intersect at a 
certain $\Phi_{\rm max}$ and the area enclosed is finite.  
The intersection point is 
\begin{equation}
\label{phimax}
\Phi_{\rm max} = {1\over 4n^2}\left({3\omega-2\over\omega}
\right)^{n/(n-2)}\left({32\pi^2\over\lambda n^3}\right)^{1/(n-2)}.
\end{equation}
Following the same trend as before, the $\sigma$ field is 
taken by quantum jumps to the largest values accessible, and 
we see from Fig.~3 that the quantum scale BoI will wedge in 
the field $\sigma$ as it approaches the intersection point, 
for which the beginning and end of inflation coincide. 
Therefore, the most typical value of the fields will be 
that of the intersection point, or strictly speaking, spread 
over its neighbourhood, as the intersection point as such 
is a region of measure zero, and the classical path 
through it has no length. The important prediction of the theories 
$n>2$, as all {\it class I} potentials, is that the crossing 
of BoI and EoI yields typical values of the fields that 
are finite. 

{\it Class I} potentials therefore predict 
a distribution of $M_P$ at the end of inflation that spans 
a range of values that is bounded above by $M_P(\Phi_{\rm
max})<\infty$. This prediction is perfectly consistent 
with our own observable universe, within a very broad 
range of parameters ($\lambda$,$\omega$,$n$) allowed by the constraint 
$M_P({\rm observed})\lsim M_P(\Phi_{\rm max})$. All finite 
values of $M_P$ satisfying this relation can take place 
with non-negligible probability, and therefore the scenario 
is a plausible one. However, {\it class II} potentials 
are problematic to reconcile with a finite value of $M_P$. 
In these cases, such as $n=1,2$, there are still regions 
that cross EoI at finite values of the fields, though their 
volume is negligible with respect to other regions for 
which the fields are arbitrarily large. Such scenarios 
are inconsistent with the observable universe on the 
basis of arguments of likelihood and can only be rescued 
by invoking very strongly the anthropic principle.

\section*{Acknowledgments}

The author is grateful to Reza Tavakol, Bernard Carr and 
Anupam Mazumdar for interesting discussions.


\end{document}